\documentclass[11pt,letterpaper,oneside]{article}
\pdfoutput=1
\usepackage{epsfig}
\usepackage{graphicx}
\usepackage{times,epsfig,makeidx,amsfonts,amsmath}
\usepackage{bm}
\usepackage{natbib}
\usepackage{lscape} 
\usepackage[plain,noend]{algorithm2e}
\usepackage{amsmath}
\usepackage{graphicx}
\usepackage{amssymb,amsmath,amsfonts,amsthm}
\usepackage{bm}
\usepackage{graphicx}
\usepackage{bbm}
\usepackage{tabularx}
\usepackage{multirow}
\usepackage{natbib}
\usepackage{amsfonts}
\usepackage{booktabs} 
\usepackage{array} 
\usepackage[above, below]{placeins}

\usepackage[a4paper,left=1in,top=1in,right=1in,bottom=1in,ignoreheadfoot]{geometry}
\usepackage{setspace}

\def\N{\mbox{N}}

\def\d{{\rm d}}

\def\half{\hbox{$1\over2$}}

\makeatletter
\renewcommand{\algocf@captiontext}[2]{#1\algocf@typo. \AlCapFnt{}#2} 
\def\@algocf@capt@plain{top}
\renewcommand{\algocf@makecaption}[2]{%
  \addtolength{\hsize}{\algomargin}%
  \sbox\@tempboxa{\algocf@captiontext{#1}{#2}}%
  \ifdim\wd\@tempboxa >\hsize
    \hskip .5\algomargin%
    \parbox[t]{\hsize}{\algocf@captiontext{#1}{#2}}
  \else%
    \global\@minipagefalse%
    \hbox to\hsize{\box\@tempboxa}
  \fi%
  \addtolength{\hsize}{-\algomargin}%
}
\makeatother

\newtheorem{lemma}{Lemma}[section]

\begin{document}



%
%
%
\title{Assigning a value to a power likelihood in a general Bayesian model}

\author{C. C. Holmes \& S. G. Walker \\ ~ \\ Department of Statistics, University of Oxford, OX1 3LB \& \\ Department of Mathematics, University of Texas at Austin, 78705. \\ cholmes@stats.ox.ac.uk \& s.g.walker@math.utexas.edu}

\maketitle

\begin{abstract}
Bayesian approaches to data analysis and machine learning are widespread and popular as they provide intuitive yet rigorous axioms for learning from data; see \cite{bernardo2001bayesian} and \cite{bishop2006}. However, this rigour comes with a caveat, that the Bayesian model is a precise reflection of Nature. 
There has been a recent trend to address potential model misspecification by raising the likelihood function to a power, primarily for {\em robustness} reasons, though not exclusively. In  this paper we provide a coherent specification of the power parameter once the Bayesian model has been specified in the absence of a perfect model.


\end{abstract}


\section{Introduction}
Bayesian inference is one of the most important scientific learning paradigms in use today. Its core principle is the use of probability to quantify all aspects of uncertainty in a statistical model, and then given data $x$, use conditional probability to update uncertainty via Bayes theorem,
$
p(\theta | x) \propto f(x; \theta) \, p(\theta),
$
where $\theta$ denotes the parameters of the model, $p(\theta | x)$ the posterior update, $f(x; \theta)$ is the data model, and $p(\theta)$ is the prior on the model parameters. Bayesian updating is {\sl coherent}, see for example \citep{lindley2000philosophy}.

The justification for Bayesian updating proceeds on an assumption that the form of the data model, $f(x; \theta)$, is correct up to the unknown parameter value $\theta$. Bayesian learning is optimal, see \cite{zellner1988optimal}, which means that posterior uncertainty is the appropriate reflection of prior uncertainty and the information provided by the data. However, this is only in the case when the model is true.
This is at odds with the scientific desire for keeping models simple in order to focus on the essential aspects of the system under investigation. 



Recently a number of papers have appeared seeking to address the mismatch and allow for Bayesian learning under model misspecification; the key reason is {\em robustness}, and the idea is to raise the likelihood to a power. See, for example, \cite{royall2003robust},
\cite{zhanga}, \cite{zhangb}, \cite{Jiang}, \cite{bissiri2013general}, \cite{walker2001}, \cite{watson2016}, \cite{miller2015}, \cite{grunwald}, \cite{syring2015}, and more generally \cite{hansen2008}.  The paper by \cite{bissiri2013general}, in particular, provides a formal motivation using a coherency principle for raising the likelihood to a power. 

For the formal Bayesian analyst, if $f(x; \theta)$ is misspecified, then there is no connection between any $\theta$ and any observation from this model, and as a consequence no meaningful prior  can be set. In this case, it is argued in \cite{bissiri2013general} that it is preferable to look at 
$-\log f(x; \theta)$ as simply a loss function linking $\theta$ and observation $x$. Then a formal general-Bayesian update of prior $p(\theta)$ to posterior $p(\theta|x)$ exists and
for the update to remain coherent it was shown that it must  be of the form,
\begin{eqnarray}
p_w(\theta | x) & \propto & f(x; \theta)^w\, p(\theta), \nonumber \\
\log p_w(\theta | x) & = & w \, \log f(x; \theta) + \log p(\theta) + \log Z_{w} \nonumber
\end{eqnarray}
where $Z_w$ is the normalising constant ensuring that the posterior distribution integrates to 1, and $w$ is a weighting parameter calibrating the two loss functions for $\theta$, namely $-\log p(\theta)$ and $-\log f(x; \theta)$ . In this way, $w > 0$ controls the learning rate of the generalised-Bayesian update, with $w=1$ returning the conventional Bayesian solution. Clearly for $w < 1$ the update gives less weight to the data relative to the prior compared to the Bayesian model, resulting in a posterior that is more diffuse, and with $w>1$ the data is given more prominence. 

The crucial question then becomes how to set $w$ in a formal manner. 
One needs to be careful as learning about $w$ can both be overdone ($w$ set too high and the posterior uncertainty is underestimated) and under done ($w$ set too low and the posterior uncertainty is overestimated). The elegant and attractive nature of Bayesian inference when the model precisely matches Nature is that the learning is achieved optimally; i.e at the correct speed. See \cite{lindley2000philosophy}, \cite{bernardo2001bayesian} and \cite{zellner1988optimal}.

In this paper we propose to set $w$ once a proper $p(\theta)$ and model $f(x; \theta)$ have been set by matching the prior expected gain in information between prior and posterior from two potential experiments; for Experiment 1 using $p_w(\theta|x)$ we compute an expected information gain between $p_w(\cdot|x)$ and $p(\cdot)$, denoted by $I_w(x)$, to be specified later. For Experiment 2 we consider the corresponding gain in information between posterior $p(\theta|x)$ and $p(\theta)$, which will be  $I_1(x)$.
Then we set $w$ so that
\begin{equation}\label{ident}
\int I_w(x)\,f_0(x)\,\d x=\int I_1(x)\, f(x; \theta_0)\,\d x,
\end{equation}
where $f_0(x)$ is the true, unknown, density and $\theta_0$ is the true parameter value if the parametric model is correct or else is the parameter value minimizing the Kullback-Leibler divergence between the true model and the parametric family of densities. So, if the model is correct, then
$f_0(x)=f(x; \theta_0)$ and $w$ will automatically be 1.
The rationale for (\ref{ident}) is coherence; that the expected gain in information for learning about $\theta_0$ from a single sample for both experiments is the same. To elaborate: Experiment 1 is assuming the data is not necessarily coming from the parametric model, the likelihood is $f(x; \theta)^w$ with prior $p(\theta)$ and $x\in f_0(x)$. According to \cite{bissiri2013general}, the $p_w(\theta|x)$ is a valid update for learning about the $\theta$ which minimizes the Kullback-Leibler divergence between $f_0(x)$ and $f(x; \theta)$; i.e. $\theta_0$, and for $w>0$ the posterior $p_w(\cdot|x)$ will be consistent for $\theta_0$ for regular models. That this is being learnt about follows from \cite{berk1966limiting}.
Experiment 2 is assuming the data is coming from the parametric model, the likelihood is $f(x; \theta)$ with prior $p(\theta)$ and $x\in f(x; \theta_0)$. Both experiments are involved with learning about the same $\theta_0$. We argue that the experimenter should be a priori indifferent between these two experiments with respect to the prior expected gain in information about $\theta_0$. Thus, $w$ is set so the prior expected gain in information is the same as that which would have been obtained if the parametric model were correct.

We can evaluate both sides of (\ref{ident}) using the observed data, $\{x_1, \ldots, x_n\}$, so the left side and right side of (\ref{ident}) are evaluated as
$$n^{-1}\sum_{i=1}^n I_w(X_i)\quad\mbox{and}\quad \int I_1(x)\, f(x; \widehat{\theta})\,\d x,$$
respectively, where $\widehat{\theta}$ is the maximum likelihood estimator. See \cite{white1982maximum} about the theory for $\widehat{\theta}$ being the appropriate estimator for $\theta_0$. In the next section we define $I_w(x)$ and in section 3 we present some illustrations.

\section {The prior expected information in an experiment} 

To quantify the prior expected information of an experiment we utilise the well established notion of Fisher information; see \cite{lehmann1998theory}.
In particular we shall consider the expected divergence in Fisher information, $F(p_1, p_2)$, between two density functions $p_1$ and $p_2$, with exact form given below; see for example \cite{otto2000generalization}.  Motivation for this choice is given in the Appendix.

The Fisher relative information divergence of a posterior update from its prior, with likelihood  $f(x; \theta)$, is given by
$$F\big\{p(\cdot), p(\cdot | x)\big\}=\int p(\theta)\,\left\{\frac{\bigtriangledown p(\theta|x)}{p(\theta|x)}-\frac{\bigtriangledown p(\theta)}{p(\theta)}\right\}^2\,\d\theta,$$
where the $\bigtriangledown$ operates on the $d$ dimensional $\theta$. This is given by
%
\begin{equation}
\label{eq:F}
F\big\{p(\cdot),p(\cdot|x)\big\}=\int p(\theta)\,\left\{\frac{\bigtriangledown f(x; \theta)}
{f(x; \theta)}\right\}^2\,\d\theta =\int p(\theta)\,\sum_{j=1}^d \left\{\frac{\partial}{\partial \theta_j} \log f(x; \theta)\right\}^2\,\d\theta.
\end{equation}
Hence, with likelihood $f(x; \theta)^w$, we have $I_w(x)=w^2\Delta(x)$, where
$\Delta(x)=F\big\{p(\cdot),p(\cdot|x)\big\}$.

This leads to

\begin{equation}
\label{eq:wset}
w =\left\{\frac{\int f(x; \theta_0)\,\Delta(x)\,\d x}{\int f_0(x)\,\Delta(x)\,\d x}\right\}^{\half}.
\end{equation}
This result also highlights why Fisher information is a convenient measure of information in the experiment as it leads to an explicit formula for the setting of $w$.

The actual setting of $w$ via (\ref{eq:wset}) is hindered by the lack of knowledge of $f_0$ and $\theta_0$. However, an empirical approach follows trivially since we can estimate $f_0(x)$ with the empirical distribution function of the data and
then estimate $\theta_0$ with $\widehat{\theta}$, the  maximum likelihood estimator. 
Thus
$$\widehat{w}= \left\{\frac{\int f(x; \widehat{\theta})\,\Delta(x)\,\d x}{n^{-1}\sum_{i=1}^n\,\Delta(X_i)}\right\}^{\half}.$$
A common simplifying choice of model would be from the class of exponential family; 
$$f(x; \theta) =\exp\left\{\sum_{j=1}^M \theta_j\,\phi_j(x)-b(\theta)\right\}$$
where the $(\phi_j(x))$ are a set of basis functions and $b(\theta)$ is the normalizing constant.
Then straightforward calculations yield
$$w^2=\frac{\int\int \sum_{j=1}^M \{\phi_j(x)-b_j'(\theta)\}^2\, f(x; \theta_0)\,p(\theta)\,\d x\,\d\theta}
{\int\int \sum_{j=1}^M \{\phi_j(x)-b_j'(\theta)\}^2\, f_0(x)\,p(\theta)\,\d x\,\d\theta},$$
where $\theta_0$ is given by 
$\int \phi_j(x)\,f_0(x)\,d x=b_j'(\theta_0)$
for all $j=1,\ldots, M$, and $b_j'(\theta)=\partial b(\theta)/\partial\theta_j$.
Hence
$$\widehat{w}^2=\frac{\int\int \sum_{j=1}^M \{\phi_j(x)-b_j'(\theta)\}^2\, f(x; \widehat{\theta})\,p(\theta)\,\d x\,\d\theta}
{n^{-1}\sum_{i=1}^n \int \sum_{j=1}^M \{\phi_j(x_i)-b_j'(\theta)\}^2\,p(\theta)\,\d\theta}.$$
In general we have, under the usual assumptions on the model that $\widehat{\theta}=\theta_0+O_p(n^{-\half})$, and that $\int \Delta^2(x)\,f_0(x)\,\d x<\infty$:

\begin{lemma}
If  $f(x; \theta_0)=f_0(x)$ then $\widehat{w}\rightarrow 1$ in probability as $n\rightarrow\infty$.
\end{lemma}

\noindent {\em{Proof}}. If we write $\gamma(\theta)=\int \Delta(x)\,f(x; \theta)\,\d x$ then we have
$\gamma(\widehat{\theta})=\gamma(\theta_0)+O_p(n^{-\half})$. 
Also, $\gamma_n=n^{-1}\sum_{i=1}^n \Delta(x_i)=\gamma(\theta_0)+O_p(n^{-\half})$ and hence we have the result as $\widehat{w}^2=\gamma(\widehat{\theta})/\gamma_n. ~~ \Box $

\section{Illustrations} We consider illustrations chosen to highlight the essential features of setting $w$,  chosen when the model is exponential family; specifically Poisson and normal.

\subsection{Poisson model}

If the model is Poisson, then for some $\theta>0$ the mass function for observation $X=x$ is given by
$f(x; \theta)=\theta^x/x!\,\,e^{-\theta}$
for $x=0,1,2,\ldots\,$. Then to find $w$ we need to evaluate the denominator and numerator in (\ref{eq:wset}),
$$D=\sum_{x=0}^\infty \Delta(x)\,f_0(x)\quad\mbox{and}\quad N=\sum_{x=0}^\infty\Delta(x)\,f(x; \theta_0)$$
where 
$$\Delta(x)=\int_0^\infty \left\{\frac{\partial f(x; \theta)/\partial\theta}{f(x; \theta)}\right\}^2\, p(\theta) \,\d \theta =\int_0^\infty (x/\theta-1)^2\, p(\theta) \,\d \theta,$$
$\theta_0$ maximizes 
$\sum_x f_0(x)\,\log f(x; \theta);$ and as $f(x; \theta)$ is Poisson we have  $\theta_0=\mu_0$ as the expected values from $f_0$, and $\sigma_0^2$ the variance from $f_0$. Hence, letting $a=\int_{\theta>0} \theta^{-2}p(\theta) \d \theta$ and $b=\int_{\theta>0} \theta^{-1} p(\theta) \d\theta$, we find, 
$D=a (\mu_0^2+\sigma_0^2)-2b\mu_0+1\quad\mbox{and}\quad N=(\mu_0^2+\mu_0)a-2b\mu_0+1.$
Then for the Poisson model fit to data arising from $f_0(x)$ we have $w^2=N/D$ and $D=a(\sigma_0^2-\mu_0)+N$.

On inspection of the result we see that when $\sigma_0^2>\mu_0$, where the data are ``overdispersed'', we find that $w<1$. The idea here is that the data will provide larger than expected observations, from a Poisson model perspective, and unless the observations are down weighted, then inference will appear overly precise. Downweighting the information in the observations will provide a more stable and practical inference for the unknown parameter.  Equally when the data are underdispersed then the Bayesian learning will be adjusted to $w > 1$ accounting for the increased precision in the data to learn about the parameter $\theta_0$ minimising the relative entropy of the model to the data distribution.

\begin{figure}[h]
\centering
\includegraphics[width=0.7\textwidth]{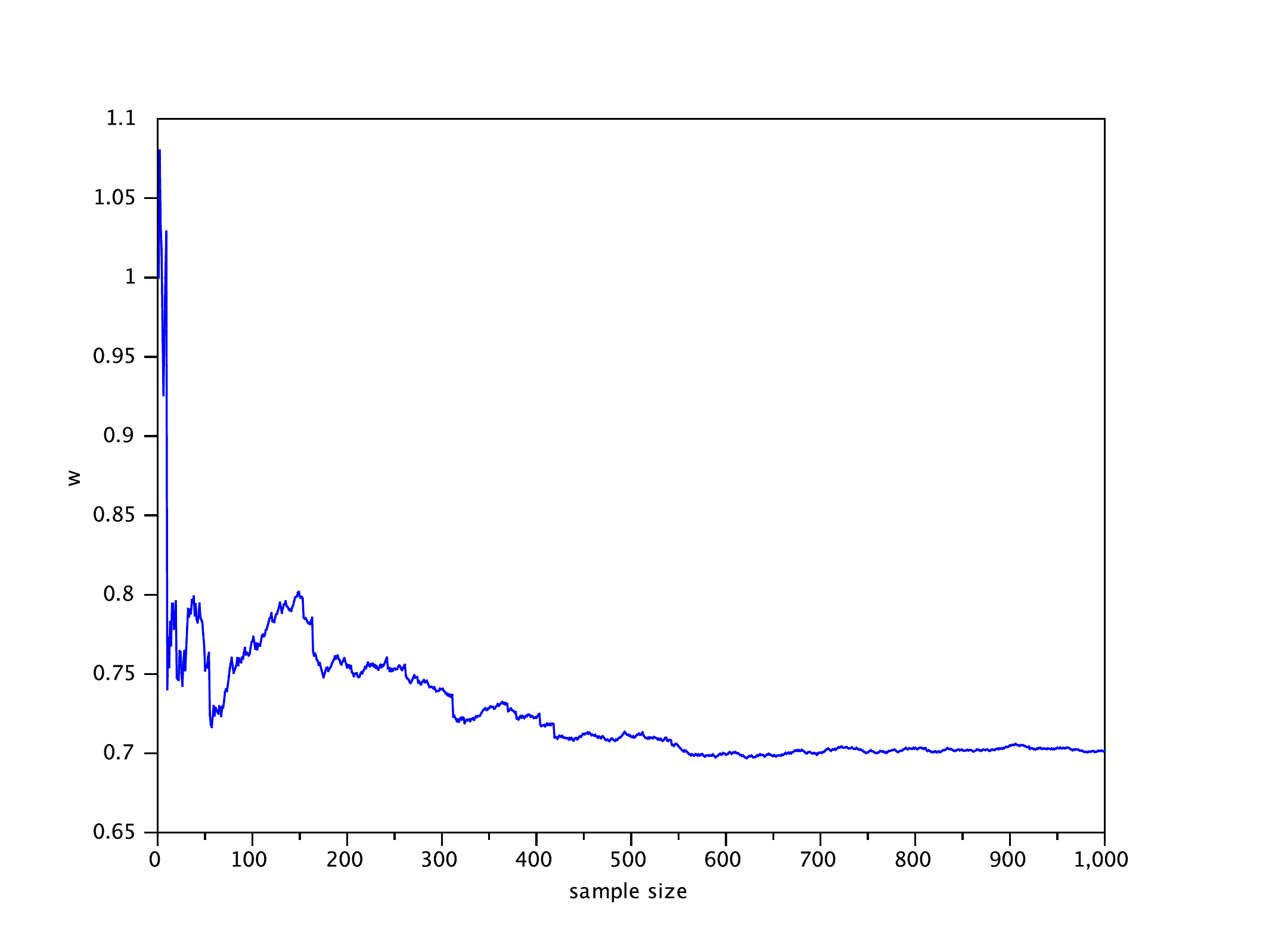}
\caption{Plot of $\widehat{w}$ against sample size: Overdispersed case, Poisson example.}
\label{fig:f1}
\end{figure}

To illustrate the performance we conducted the following experiment. We took $n=1000$ observations
from an overdispersed model, so $X$ given $\phi$ is Poisson with mean $\phi$ and $\phi$ is from the gamma distribution with mean $3.33$ and variance  $11.11$. Thus the variance of the data is
$14.44$ while the mean of the data is $3.33$, so there is a substantial amount of overdispersion. 
The prior for $\theta$ in the Poisson $f(x; \theta)$ model was taken to be gamma with mean $3$ and variance $3$. 
For this experiment we then computed $\widehat{w}$ using the sample mean $(\bar{x})$ and sample variance $(S^2)$; 
$\widehat{D}=a(\bar{x}^2+S^2)-2b\bar{x}+1\quad\mbox{and}\quad \widehat{N}=a(\bar{x}^2+\bar{x})-2b\bar{x}+1.$
Thus
$$\widehat{w}^2=\frac{a(\bar{x}^2+\bar{x})-2b\bar{x}+1}{a(\bar{x}^2+S^2)-2b\bar{x}+1}.$$
We plot the $\widehat{w}$ against sample size in Fig~\ref{fig:f1}, and note that essentially the $\widehat{w}<1$, with convergence to a number lower than 1.

On the other hand, if the model was true (the so called $M$-closed perspective in \cite{bernardo2001bayesian}), then $S^2-\bar{x}\rightarrow 0$, then $\widehat{w}^2\rightarrow 1$. Moreover, using standard asymptotic, large sample size $n$, properties of models and estimators, we have that $1-\widehat{w}^2\rightarrow 0$ at a speed of $n^{-\half}$. 


\subsection{Exponential family} We provide some further analysis of the general case for the exponential family based on
$f(x; \theta)=c(x)\exp\{\theta x-b(\theta)\}.$
Then following the same strategy as in the previous sub-section, 
and using (\ref{eq:wset}), 
where now
$\Delta(x)=\int \{x-b'(\theta)\}^2\,p(\theta)\,\d\theta$
and $b'(\theta_0)=\int x f_0(x)\,d x=\int x p_{\theta_0}(x)\,d x$, we can show
$$w^2=\frac{b''(\theta_0)+\int \big\{b'(\theta_0)-b'(\theta)\big\}^2p(\theta)\,d\theta}
{\sigma_0^2+\int \big\{b'(\theta_0)-b'(\theta)\big\}^2p(\theta)\,\d\theta}$$
which is estimated via
$$\widehat{w}^2=\frac{b''(\widehat{\theta})+\int \big\{\bar{X}-b'(\theta)\big\}^2p(\theta)\,\d\theta}
{S^2+\int \big\{\bar{X}-b'(\theta)\big\}^2p(\theta)\,\d\theta}.$$
Thus, $w$ will converge to 1 or otherwise depending on how the sample variance $S^2$ compares with the variance estimator from the model; namely $b''(\widehat{\theta})$.
Even in the case of regression models, the basic idea is the same when $\Delta(\cdot)$ is quadratic, as it would be for example in the case of a normal linear regression model.

\subsection{Normal model} Here we consider a normal model with unknown mean $\theta$ and variance 1. The prior for $\theta$ is normal with mean 0 and precision parameter $\lambda$. The aim here is to compare our selection of $w$ with an alternative using the Kullback-Leibler divergence; i.e.
to set $w$ based on matching
$$\int D\{p_w(\cdot|x),p(\cdot)\}\,\d F_n(x)=\int D\{p(\cdot|x),p(\cdot)\}\,f(x;\widehat{\theta})\,\d x,$$
where $D(q,p)=\int q\log(q/p)$. Although there is no closed form solution for $w$ here, we can evaluate it numerically.

First we considered the overdispersed case and so generated 50 observations from a normal distribution with precision 0.2 and use the prior for $\theta$ to have mean 0 and precision 0.01. Then we looked at the underdispersed case and generated 50 observations from a normal distribution with precision 4 and again use the prior for $\theta$ to have mean 0 and precision 0.01

\begin{figure}[h]
\centering
\includegraphics[width=0.9\textwidth]{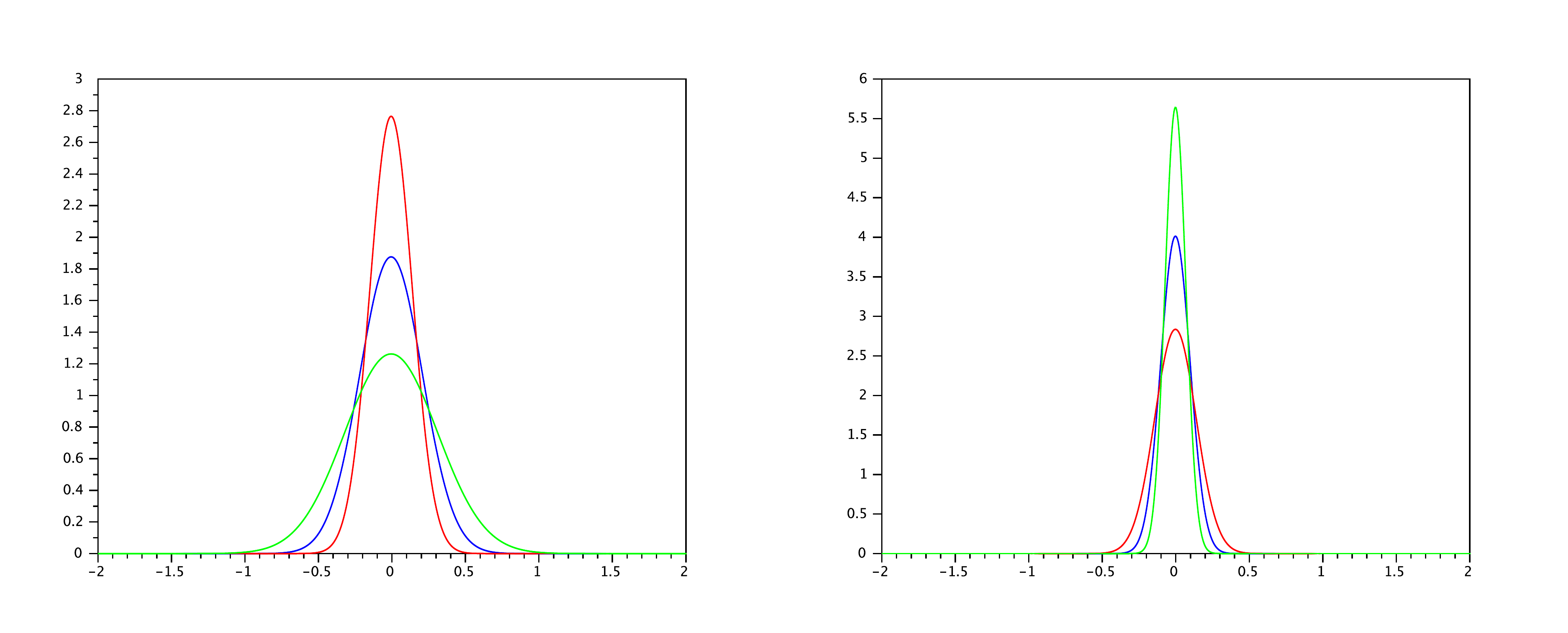}
\caption{Posterior distributions in the overdispered case (left figure) and the underdispersed case (right figure) for normal example: posterior based on Fisher distance $w$ in blue; posterior based on Kullback-Leibler $w$ in red; and true posterior using the correct model from which data are generated in green.}
\label{fig:f2}
\end{figure}

In Fig \ref{fig:f2}, on the left side,  we plot three posterior distributions: blue is the posterior using the $w$ from our Fisher information distance; red is the posterior using the $w$ obtained from the Kullback-Leibler divergence, and the green is the correct  posterior had the model been used with the correct precision parameter of 0.2.

On the right side of Fig \ref{fig:f2} we again plot three posterior distributions: blue is the posterior using the $w$ from our Fisher information distance; red is the posterior using the $w$ obtained from the Kullback-Leibler divergence, and the green is the correct  posterior had the model been used with the correct precision parameter of 4. In both cases we see that our posterior is closer to the posterior based on the correct model; i.e. replacing 1 with the precisions 0.2 and 4, respectively.

\section{Discussion} It can be argued that all models are misspecified. Under such a scenario there is no formal connection  between  any observed $x$ and any $\theta$ when looking at $f(x; \theta)$ as a density function. On the other hand, when viewed as a loss function, $-\log f(x; \theta)$, and learning about $\theta_0 = \arg \min_{\theta \in \Theta} \int f(x; \theta) f_0(x)\,\d x$, we can interpret the correspondence between $x$ and the object of inference $\theta$. However, as pointed out in \cite{bissiri2013general},  in this setting there is a free parameter $w$ introduced by the model misspecification. In this paper we have introduced principles for the specification of $w$ which provides an a priori coherent agenda in terms of prior expected gain in information about $\theta_0$.

\section*{Appendix: Motivation for Fisher information distance}  As shown in \cite{walker2016}, the expected (with respect to the prior predictive) Fisher information distance between prior and posterior is given by
\begin{equation}\label{fishi}
\int \bar{p}(x)\,F(p(\cdot|x),p(\cdot))\,\d x=\int J(\theta)\,p(\theta)\,d\theta=E\{J(\Theta|X)\}-J(\Theta)
\end{equation}
where $J(\theta)$ is the Fisher information for $\theta$, $\bar{p}(x)$ is the prior predictive $\bar{p}(x) = \int f(x; \theta) p(\theta) d \theta$, and 
$J(\Theta)=\int p'(\theta)^2/p(\theta)\,d\theta$ is known as the Fisher information for the density $p(\theta)$, while $J(\Theta|X)$ is the Fisher information for the posterior given $X$. So it has similar properties to the Kullback-Leibler divergence which relies on expected differential entropy between prior and posterior.

However, instead of  using 
$$F\{p(\cdot|x),p(\cdot)\}=\int p(\theta|x)\,\left\{\frac{\partial}{\partial\theta}\log \frac{p(\theta|x)}{p(\theta)}\right\}^2\,\d \theta$$
to get (\ref{fishi}), we use
$$F\{p(\cdot),p(\cdot|x)\}=\int p(\theta)\,\left\{\frac{\partial}{\partial\theta}\log \frac{p(\theta|x)}{p(\theta)}\right\}^2\,\d \theta.$$
For the former is suited to the idealized setting of a correct model; whereas we are trying to evaluate the prior and posterior discrepancy, i.e. 
$$\left\{\frac{p'(\theta|x)}{p(\theta|x)}-\frac{p'(\theta)}{p(\theta)}\right\}^2=\left\{\frac{\partial}{\partial\theta}\log f(x;\theta)\right\}^2=S^2(x,\theta),$$
where $S(x,\theta)$ is the usual score function, with respect to prior beliefs, for it is only the prior beliefs we assume common to both experimenters; i.e. the one using $I_1$ and the one using $I_w$.

We can elaborate further: the prior expected Fisher information; i.e. $E_{p(\theta)}\{J(\theta)\}$, is
$$\int J(\theta)\,p(\theta)\,\d\theta=\int F\{p(\cdot|x),p(\cdot)\}\,\bar{p}(x)\,\d x=\int\int S^2(x,\theta)\,p(\theta)\,f(x;\theta)\,d\theta\,\d x.$$
This would be the expected information in a single sample as an expected discrepancy between prior and posterior.
However, this expected Fisher information is provided under the idealized setting that the joint density of $(x,\theta)$ for the expectation of $S^2(x,\theta)$ is $p(\theta)\,f(x;\theta).$
It would be unrealistic for us to assume the marginal density for $x$ is $\bar{p}(x)$, even for the Bayesian assuming $f(x;\theta)$ is correct. A more realistic estimation of the expected squared score function, i.e. information in a single sample,  would be to use the empirically determined joint density $p(\theta)\,f(x;\widehat{\theta})$.  

For the Bayesian using $f(x;\theta)^w$, the score function is $S_w(x,\theta)=w\,S(x,\theta)$, and so would estimate the information, using the product measure of the prior and empirical distribution function, $F_n(x)$, since this Bayesian is assuming the model incorrect. 
Matching these two forms of information from a single sample and about the same parameter, we have 
$$\int \int S_w^2(x,\theta)\,p(\theta)\,\d\theta\,d F_n(x)=\int \int S^2(x,\theta)\,p(\theta)\,f(x,\widehat{\theta})\,d\theta\,d x$$
where the term on the left is given by
$w^2 \int \int S^2(x,\theta)\,p(\theta)\,\d\theta\,d F_n(x)$ and recall that $\int S^2(x,\theta)p(\theta)\,\d\theta=F\{p(\cdot),p(\cdot|x)\}$. In short, we are using the square of the score function as a measure of information in a single sample which also has the interpretation in terms of Fisher distance between prior and posterior. 

\section*{Acknowledgements} The authors are grateful to two anonymous referees and an Associate Editor for comments and suggestions on a previous version of the paper.


\newpage

\end{document}